\documentclass[twocolumn,aps,pre]{revtex4}

\usepackage{graphicx}
\usepackage{dcolumn}
\usepackage{bm}
\usepackage{hyperref}%

\begin{document}


\title{Complex light: Dynamic phase transitions of a light beam in a nonlinear
non-local disordered medium}

\author{Claudio Conti}
\email{claudio.conti@phys.uniroma1.it}
\homepage{http://nlo.phys.uniroma1.it}
\affiliation{
Research center ``Enrico Fermi'' Via Panisperna 85/A 00184, Rome, Italy and \\
Reserch center SOFT INFM-CNR, University ``La Sapienza,''  P. A. Moro 2, 00185, Rome, Italy
}%

\date{\today}

\begin{abstract}
The dynamics of several light filaments (spatial optical solitons) propagating
in an optically nonlinear and non-local random medium is investigated using the
paradigms of the physics of complexity. Cluster formation is interpreted as a dynamic phase
transition. A connection with the random matrices approach
for explaining the vibrational spectra of an ensemble of solitons is
pointed out. 
General arguments based on a Brownian dynamics model are validated by the numerical
simulation of a stochastic partial differential equation system.
The results are also relevant for Bose condensed gases and plasma physics.
\end{abstract}

\maketitle

\section{Introduction}
At low temperature, the dynamics of complex media is dominated by
the  potential  energy  landscape  (PEL), i.e.  the  multi-dimensional
surface  of  the potential  energy  as  a  function of  the  molecular
coordinates.  \cite{Goldstein69,Stillinger82,Jonsson88,Sastry98} 
A  disordered  system   sampling  different  PEL
configurations undergoes  ``dynamic phase transitions,''  perhaps one of
the most spectacular ideas  of the physics of complexity 
\cite{Bhattacharya99,Angelani00,Broderix00,Grigera03}.
For the dynamic  glassy transition, observed in a  certain class of
(glass-forming) supercooled liquids, this ``configurational sampling''
is at  the origin  of an  increase of viscosity  of several  orders of
magnitudes, up to values  comparable to solids 
(for recent reviews see for example \cite{PhysicaD,Debenedetti01}).
This phenomenon
cannot  be directly  ascribed  to a  purely thermodynamic  transition;
conversely,  it is  now widely  accepted as  a dynamic  effect. During
cooling, different PEL regions are visited, and the  almost abrupt change
of viscosity  is associated to the transition from ``saddle-dominated'' to
``minima-dominated''  PEL  basins.  Since  phonons  are  the  elementary
excitations  around  minima of  the  PEL,  this  process is  
also denoted ``phonon-saddle'' transition,  and can be realized
while keeping fixed the temperature and acting on some parameter, like
the particle density or the interaction range.  \cite{Parisi03} 

It  is  not difficult  to recognize  the
fundamental character of  these ideas, and the fact  that they are not
limited to  the specific contexts where they  originally developed. 
In this article I will show that
a light beam propagating in disordered medium (i.e. a medium 
with negligible optical losses whose
refractive index is randomly varying)
may undergo a sort of dynamic phase transition.
This happens when, 
due to an optically nonlinear response, multiple filaments are generated.
 Their number and properties depend on the mutual
interaction range and, in essence, they behaves like molecules in 
a complex medium,  
exhibiting dynamic phases. The process can described in terms
of the appropriately defined ``inherent structures'' and ``saddles,'' in
perfect analogy with the physics of disordered materials. 
This qualifies as a sort of \textit{soft}, or \textit{complex}, light. 
 
Here the  word ``filament'' (roughly) identifies a
spatial soliton (SS),  which is a non-diffracting light beam  generated in an
optically nonlinear medium,  with an  intensity  dependent refractive
index.  \cite{BoardmanBook,TrilloBook, KivsharBook} 
If a  sufficiently intense  laser light  propagates in
such  a  material,   self-induced  trapping  counteracts  the  natural
tendency to  diffract, and a tightly  focused SS can  be observed.  
For example, in nematic liquid crystals (NLC), with laser wavelength in the
near infrared, 
it is possible to  generate  very thin  (few  microns  waist)  SSs, able  to  propagate
undistorted   for  millimeters:  hundreds   of  times   the  distances
attainable in absence  of a nonlinear self-action. \cite{Assanto03,Peccianti04nature,Hutsebaut05}  

At each SS is associated an optically induced  perturbation  $\Delta n$ to the refractive index,
which is at the origin of  the self-focusing.  If the incident beam is
sufficiently wide many SSs are generated  by the
same input, as shown for example in \cite{Peccianti04nature,Peccianti03,Peccianti03b}.  
These filaments may propagate with
various degrees of interaction, relying on the material properties
and, in particular, on the so-called ``non-locality'', which can be kept
in mind as  the ratio between the spatial  extension of the refractive
index perturbation induced by one filament and its transverse intensity
waist.  
In NLC, the degree of nonlocality can be simply controlled 
by a voltage bias,\cite{Peccianti05} and hence the dynamic phase transition may
be induced accordingly.\cite{Conti05submitted}

The following results  not only are well suited to describe 
light-soft-matter interaction (as in
\cite{Peccianti03,Peccianti04nature, Conti05submitted,Peccianti05,Conti05}), 
but can be applied to interpret the dynamics of ultrashort laser
pulses propagating in air, \cite{Skupin04,Rodriguez04} as well
solitons in photorefractives,\cite{Segev92,Chen02} media with thermal nonlinearities and plasmas,
\cite{Henninot02,Yakimenko04,Litvak75} semiconductors
\cite{Ultanir03}, 
discrete solitons (see for example \cite{Fleischer03} and references therein), plasmas \cite{Litvak75} and	
Bose-condensed gases \cite{Khaykovich02,Strecker02}.

This article is aimed to the introduction of the leading idea, and to a
phenomenological description, supported by Brownian dynamics and stochastic partial differential equations (PDEs) numerical simulations.
In order to simplify as much as possible the presentation no attempt
will be made to a theoretical analysis, which is deferred to future publications.
In section II the model for the propagation of several light filaments
in a random medium is linked with Brownian dynamics.
In section III evidences of phase transitions from
numerical simulations are reported.
Section IV is dedicated to the definition and the analysis of the 
so-called inherent structures.
In section V the generalized inherent structures are considered for
the final settlement of the dynamic phase transition.
In section VI the noise quenching process is addressed. 
In section VII a connection with the theory of random matrices and the
so-called dynamic structure factor is established.
In section VIII numerical simulations of a well known model in 
nonlocal soliton theory validate the general arguments of the manuscript.
Conclusions are drawn in section IX.
\section{Reduction to a Brownian dynamics model}
For the sake of simplicity, the analysis is done with reference to
 one-dimensional
(1+1D propagation) beams.
Indeed, differently from standard thermodynamic transitions, dynamic transitions can be obtained 
in low dimensional systems. Additionally, the considered case reflects typical experimental geometries
for the investigation of modulational instability (see e.g. \cite{Peccianti03,Peccianti03b}).
The generalization to higher dimensional problems of what follow can be readily imagined.
Consider the Fock-Leontovich equation, which describes the 
paraxial optical propagation in a nonlinear medium:
\begin{equation}
\label{Foch}
2ik\frac{\partial A}{\partial z}+\frac{\partial^2 A}{\partial x^2}+2k^2 \frac{\Delta n[I]}{n}A=0\text{.}
\end{equation}
$A(x,z)$   is the  complex amplitude of the optical field such that 
$I=|A(x,z)|^2$ is the intensity, $k=2\pi n/\lambda$ is the wave-vector and $n$ the refractive index 
at wavelength $\lambda$ in absence of non-linear effects. $\Delta n$
is the optical induced perturbation
to $n$, its
functional relation  with $I$ is non-local:  the specific distribution
of $\Delta n(x,z)$ 
depends on the whole profile $I(x,z)$. 
Various models are given in the literatures relating $\Delta n$ and
$I$, as the Kukhtarev equations for photorefractives, (for a review, see Del Re
and coworkers in \cite{TrilloBook}) the 
heat equation for thermal nonlinearities, \cite{Litvak75,Yakimenko04}
re-orientational equations for liquid crystals, \cite{Conti03} mode coupling theory for soft-matter, \cite{Conti05} or generic
nonlocalities. \cite{Snyder97,Perez00,Bang02,Krolikowski04}

For relatively low intensities $\Delta n$
is linear with $I$ and, in the presence of many incoherent filaments,
 it can  be written as the  sum of the intensity  distributions of the
 filaments, with $p=1,2,...,N$.  
The overall $\Delta n$ is thus the sum of the index perturbations of each  SS:
\begin{equation}
\label{deltan}
\Delta n[I]=\displaystyle\sum_{p=1}^{N}\Delta n[I_p]\text{.}
\end{equation}
Eq. (\ref{deltan}) is valid whenever $\Delta n/n<<1$, as typically verified  in the reported experiments, and when
     the filaments  are mutually  incoherent, i.e. the  relative phase
     between pairs of them  is randomly varying. The latter hypothesis
     typically holds  in soft-matter, by taking into  account that the
     SSs spontaneously generate from noisy intensity perturbations, and
     propagate in a thermally  fluctuating medium. Furthermore, in the
     presence  of many filaments  the effects  of the  relative phases
     and, more in general, of the specific form of $\Delta n[I]$ can be
    negligible.  
This is analogous to the typical approach in statistical physics
     where, 
quite often, the particular profile of
     pair-interaction potentials can be replaced by some simple model like the Lennard-Jones potential.  \cite{SimpleLiquids}

Following  a  perturbative analysis, the effect of the  index perturbation due to all SSs
     on the trajectory of the  generic filament $p$ is considered. 
For $N$ identical {\it stable} solitons with intensity bell-shaped profile
$I_{S}(x)$, $\Delta n_{S}(x)=\Delta n[I_{S}](x)$, and average
 position  $x_p(z)$, using  the  Ehrenfest's  theorem  of
     standard  quantum mechanics,  applied to  the Schr\"odinger-like
     Eq. (\ref{Foch}), one has for the generic filament
\begin{equation}
\label{particle0}
m\frac{d^2 x_p}{d z^2}=-\displaystyle\int_{-\infty}^{\infty} I_{S}(x-x_p)
\frac{\partial \Delta n/n}{\partial x}dx\text{,}
\end{equation}
with $m=\int I_S(x) dx$ the power (per unit length along the $y$ direction)
into each filament, which plays the role of the particle mass.
The index perturbation is 
\begin{equation}
\label{deltanlong}
\frac{\Delta n}{n}=\frac{1}{n}\displaystyle\sum_{q=1}^N \Delta n_{S}(x-x_q)\text{,}
\end{equation}
which used in (\ref{particle0}) yields
\begin{equation}
\label{particle1}
\begin{array}{l}
m\displaystyle\frac{d^2 x_p}{d z^2}=\\
\\
\displaystyle\sum_{q=1}^{N}\int_{-\infty}^{\infty} I_{S}(x-x_p)
\frac{\partial \Delta n_S(x-x_q)/n}{\partial x}dx=\\
\\
-\displaystyle\sum_{q=1}^{N}\int_{-\infty}^{\infty} 
\frac{\partial I_{S}}{\partial x}(x-x_p)
\frac{ \Delta n_S(x-x_q)}{n}dx=\\
\\
\displaystyle\frac{\partial}{\partial x_p}\displaystyle\sum_{q=1}^{N}\int_{-\infty}^{\infty} 
 I_{S}(x-x_p)
\frac{ \Delta n_S(x-x_q)}{n}dx=\\
\\
-\displaystyle\frac{\partial}{\partial x_p}\displaystyle\sum_{q=1}^{N}
 V(x_p-x_q)
\end{array}
\end{equation}
with
\begin{equation}
V(x)=-\frac{1}{n}\int_{-\infty}^{\infty} \Delta n_S(\xi+\frac{x}{2})I_S(\xi-\frac{x}{2})d\xi\text{,}
\end{equation}
the pair-interaction potential. 
Many derivations of similar results can be found in the literature on
solitons and solitary waves  (see
e.g. \cite{PerezGarcia03,Crasovan03} 
and references therein). Here the analysis is specialized for 
the potential energy landscape interpretation.

In (\ref{particle1}) the self-interaction  term ($p=q$) can be retained
since it clearly gives a vanishing contribution. Finally
\begin{equation}
\label{particle2}
m\frac{d^2 x_p}{d z^2}=-\frac{\partial \Phi}{\partial x_p} 
\end{equation}
and $\Phi=\Phi(x_1,x_2,...,x_N)$ is the overall potential energy
surface (the PEL) given by the sum of pair-wise
interaction terms: 
\begin{equation}
\label{PEL}
\Phi=\frac{1}{2}\displaystyle\sum_{j=1}^{N}\sum_{k=1}^{N}V(x_j-x_k)\text{.}
\end{equation} 
The dynamics along
     the  direction of  propagation is  hence formally  reduced  to an
     ensemble of particles, evolving with ``time'' $z$. The  fluctuations  of  the  medium  result into  a  random
contribution to  $\Delta n$ that  can be phenomenologically included  in the model  as a
Langevin force $\eta_p(z)$:
\begin{equation}
\label{particle3}
m\frac{d^2 x_p}{d z^2}=-\frac{\partial  \Phi}{\partial x_p}+\eta_p(z)\text{.}
\end{equation}
In the following, I will take for $\eta_p$ a normally distributed white
noise:
\begin{equation}
<\eta_p(z)\eta_q(z')>=S_p^2 \delta_{pq}\delta(z-z')
\end{equation}
with $S_p^2$ the noise ``power'' and the brackets denoting a statistical
average over disorder. 
$\eta_p(z)$ 
takes into account the fluctuations of the refractive index and defines
     a realization of the random soft-medium. 

Note that, in typical Langevin models, the random term is accompanied by a
dissipative term, which is in general dependent on the lossy
mechanisms in the medium, like viscosity. 
In the limit of small losses and small noise such a term can 
be neglected, as it will be done in the following in order to leave 
 the treatment as general and
simple as possible. The qualitative agreement with numerical results in section VIII, and 
experiments \cite{Conti05submitted}, supports this approach.

The explicit shapes of $I_S$ and $\Delta n_S$ are due to the
 particular  nonlinear  mechanism;   for  the  present  purpose  a
     Gaussian ansatz for both of them is appropriate, since non-local 
optically nonlinear media are being considered. \cite{Snyder97,Bang02,Conti03}
Taking 
\begin{equation}
\begin{array}{l}
I_S(x)=I_0 \exp\left(-\displaystyle\frac{x^2}{2 w^2}\right)\\
\\
\Delta n_S(x)=\Delta n_0 \exp\left(-\displaystyle\frac{x^2}{2 v^2}\right)
\end{array}
\end{equation}
gives
\begin{equation}
\label{pairwisepotential}
V(x)=V_0\left[1-\exp\left(-\frac{x^2}{2 u^2}\right)\right]
\end{equation}
with $V_0=\Delta n_0 I_0[2 \pi v^2 w^2/(v^2+w^2)]^{1/2}$ and
$u^2=v^2+w^2$.
$V(x)$ is a Gaussian function within an
arbitrary additive constant. It is written as in
(\ref{pairwisepotential}) in order to have a vanishing $\Phi$ when
all the solitons are in the same position (``condensed phase'').
     $u^2$ is the sum of the  variances of $I_S$ and $\Delta n_S$ and provides a measure
     of the interaction range  between the SSs for a fixed $w$, like
$u/w$ that will be used in the following.
Each filament increases the  refractive index 
(the medium is assumed to be focusing, $\Delta n_0>0$) and the
     interaction is  purely attractive, so that $V_0>0$.  

The previous formulation points
     the  connections  with statistical  physics  while reducing  the
     model to  a system of interacting  classical particles undergoing
     Brownian  motion,  a  typical   model  for  colloids,  where  the
     formation  of clusters  at the  glass-transition is  a well-known
     process (see e.g.  \cite{Donati99,Likos01,Weeks04}).

However, before proceeding, it is fruitful to point out a subtle issue
     associated to Eqs. (\ref{particle3}) and (\ref{pairwisepotential}).
It is well expected
     that a finite number of classical particles, interacting by a
     purely attractive potential, will oscillate around the center of
     mass at equilibrium (i.e. for long times). This implies that 
only one energy minimum does exist and corresponds to the condensed
     phase. At a first glance, no dynamic phase transition, due to local minima of the
     PEL, is expected.
Nevertheless, in the case under consideration, it is clearly not possible to
consider an arbitrarily large propagation distance (corresponding to
     long times). If losses
are negligible and if the maximum observation range (in $z$) is
limited by the spatial extension of the sample, damping mechanisms can
be neglected and local PEL minima play a role. 

\section{Numerical simulations of the Brownian dynamics model}
A typical distance
between resolvable (by means of the observation of scattered light from the
sample) spatial solitons, generated for example by modulational
instability, can be taken as $6$ times their waist $w$. 
Hence, considering thin solitons with waist $w=5\mu m$ and taking tens of
filaments implies an input waist of the order of $500\mu m$, which is
comparable to those typically employed in experiments (see
e.g. \cite{Peccianti03,Peccianti04nature}).
In the following two representative cases will be considered:
$N=10$ and  $N=30$ (the principal dynamic phase transition, considered below,
is obtained up to the largest considered $N=100$,
not reported).
	In all the simulations the filaments are chosen uniformly
	distributed at $z=0$, with mutual distance $6w$.

 The noise ``power'' is measured by the
     corresponding adimensional quantity
\begin{equation}
\nu_p^2=\frac{w}{V_0^{3/2} m^{1/2}}S_p^2\text{,}
\end{equation}
which is taken independent from $p$, for the sake of simplicity:
$\nu_p^2=\nu^2$.
The amount of noise to be included in the simulations clearly relies
on the specific material (in particular, on the thermal coefficient of
the refractive index and on the sample temperature); 
however it is found that the numerical results
are very robust with respect to noise, and very similar findings are 
obtained when $\nu$ is varied by order of magnitudes.
Hence only the case $\nu=0.001$ will be reported 
as a representative example.
The stochastic ordinary differential equations (\ref{particle3}) are solved by
a second order scheme, whose accuracy has been thoroughly investigated and compared 
with other approaches.\cite{qiang00}
The results have been validated by halving the integration step and doubling the number
of realization in many cases.

\begin{figure}
\includegraphics[width=6cm]{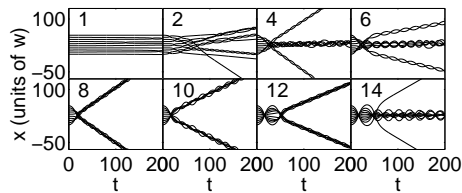}
\caption{Filaments trajectories Vs the normalized propagation
coordinate $t$ for a given noise realization
and various values of the interaction range $u/w$ (here $N=10$). \label{figN10}}
\end{figure}

\begin{figure}
\includegraphics[width=6cm]{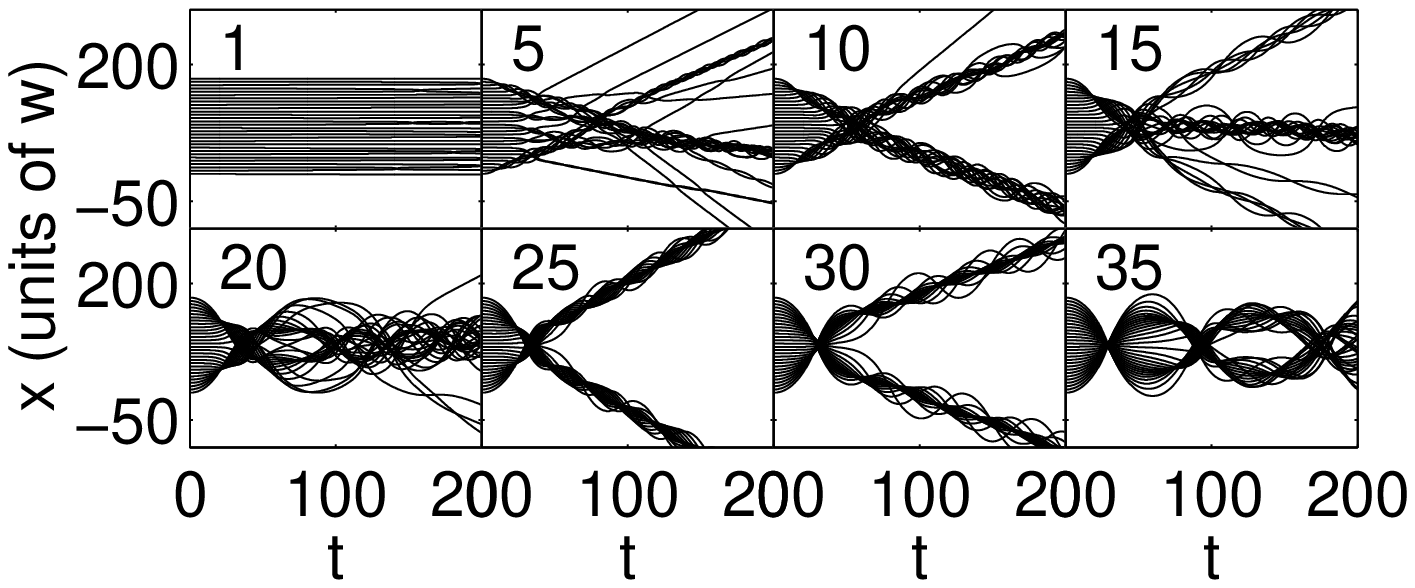}
\caption{
Filaments trajectories Vs the normalized propagation
coordinate $t$ for a given noise realization
and various values of the interaction range $u/w$ (here $N=30$).\label{figN30}}
\end{figure}

 As discussed above,
a dynamic transition is attained  while increasing the
     density, or  equivalently the interaction  length
     $u/w$.\cite{Parisi03}  
Here, this corresponds  to increase  the degree  of nonlocality.
In figure  \ref{figN10}, some realizations  
of the SSs trajectories, obtained by  the numerical
     solution  of Eqs.  (\ref{particle3}) when $N=10$, are shown for various $u/w$.
The  adimensional ``time'' $t=z/[w(V_0/m)^{1/2}]$ is  used on the horizontal axis.  
For small $u/w$ the filaments propagate in the presence of a reduced interaction.
Conversely, while increasing $u/w$, various clusters are formed and their number
and positions vary with each realization of the noise.
A similar result is obtained in the case $N=30$ (fig. \ref{figN30})

In figures \ref{figmeanposn10} and \ref{figmeanposn30},
the ``final'' (i.e. at a fixed $t=t_{max}$) 
position of each SS is shown Vs $u/w$, with  
the results for $10$ noise realizations superimposed.
Clearly, in certain ranges of the control parameter $u/w$
the statistics of the final positions are highly peaked around
$2$ or $3$ clusters, while they spread over a broad region in other ranges. 
The appearance of an interval for $u/w$ where two dominant
clusters are generated is evident. This is referred to as
the ``principal dynamics phase transition''. In the case $N=10$ a
phase with $3$ clusters is also present and it is
somehow more noisy in the case for $N=30$.
For a very large $u/w$ the nonlocality is such that all the SSs
oscillate around an equilibrium position, this has been above indicated 
as the condensed phase.

Similar results are obtained for an odd number of 
filaments (e.g. $N=21$). In that case, in correspondence
of the principal dynamic phase transition, an additional SS is found at
the middle of the two clusters.
 
An open issue is the existence of 
fractal structures as those investigated in \cite{Dmitriev02}.
\section{The inherent structure}
\begin{figure}
\includegraphics[width=6cm]{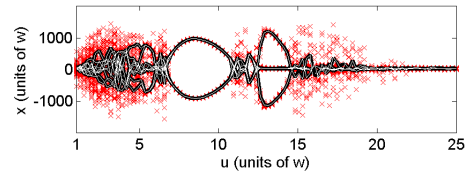}
\caption{(Color online) Crosses, filament positions at $t_{max}=1000$
for $10$ noise realizations; thick black
line, average position (see text); white line, inherent
structures (here $N=10$). 
\label{figmeanposn10}}
\end{figure}
\begin{figure}
\includegraphics[width=6cm]{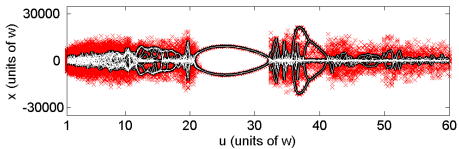}
\caption{(Color online) Crosses, filament positions at $t_{max}=6000$
for $10$ noise realization; thick black
line, average position (see text); white line, inherent
structures (here $N=30$). 
\label{figmeanposn30}}
\end{figure}

Figures \ref{figmeanposn10} and \ref{figmeanposn30} also  show the
average filament positions Vs $u/w$, when $N=10$ and $N=30$ respectively.
Note that the positions are determined at a very large $t$ so
that the clusters are ``stabilized,''
and that the average positions for each filament (thick black line in the figures)
are shown superimposed (i.e. they are not the average positions among all the filaments), so that the thick black line provides
a visualization of cluster distribution.

Disordered phases are alternated with others in which a fixed
number of clusters is obtained. 
In order to address the existence of some kind of phase transition it
is necessary to introduce a ``control parameter'', as outlined in the 
mentioned literature. \cite{Grigera03,Debenedetti01}
With  this aim,  I start pointing  out the \textit{inherent structure}  (IS) associated to
     the  numerical simulations.  Once fixed  a maximum  value  for the
     time $t_{max}$, the final distribution of  filaments is used as guess for a
     conjugate gradient minimization  procedure that finds the nearest
     minimum of the interaction potential $\Phi$. 
The corresponding vector of positions $(x_1,x_2,...,x_N)$ is the IS.  \cite{Stillinger82,Stillinger83}
Its role  is evident  when superimposing  the plots  of  the average final
     positions 
for the considered realizations (thick black line in Figs. \ref{figmeanposn10} and \ref{figmeanposn30})  and those
of the average IS  (thin  white  line in Figs. \ref{figmeanposn10} and \ref{figmeanposn30}). 
Clearly,  the latter provide information on  the number and
     the  positions  of  the   generated  clusters.  
\begin{figure}
\includegraphics[width=8.3cm]{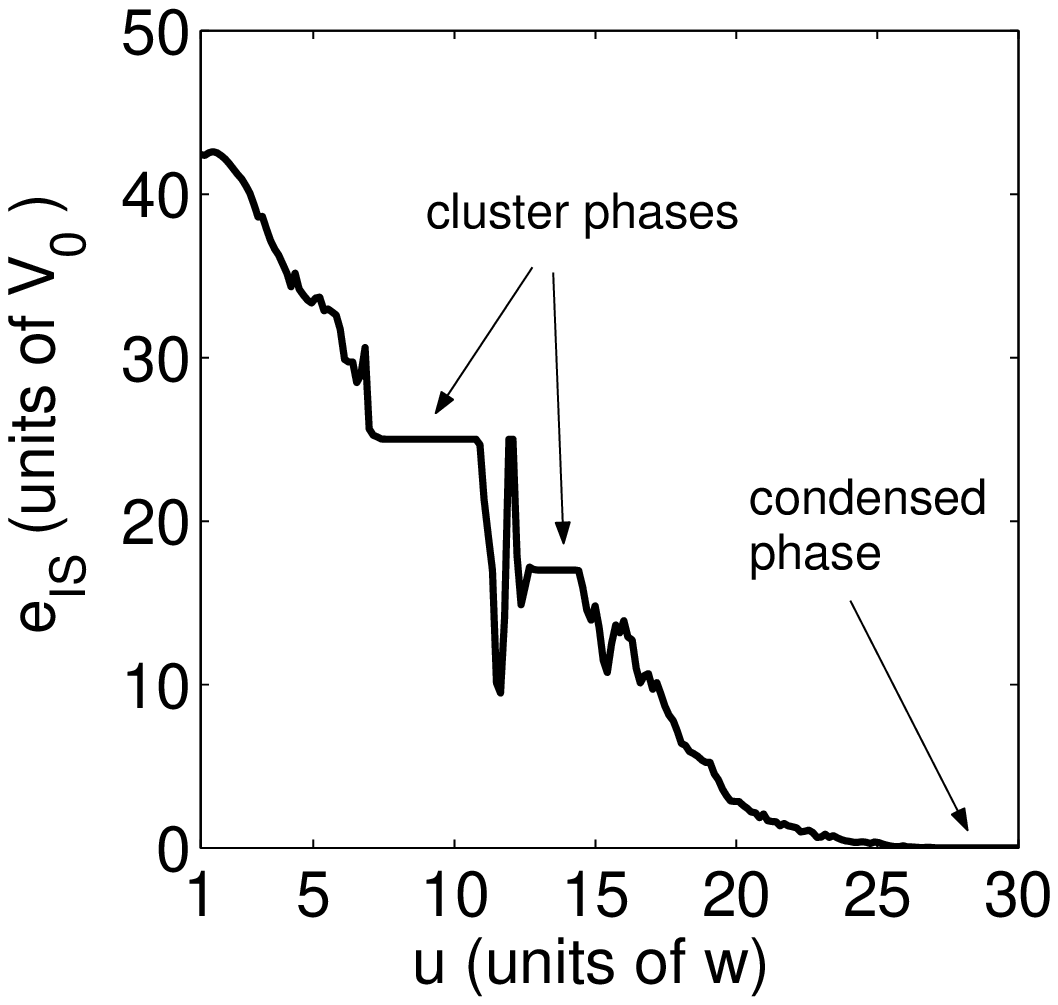}
\caption{Average potential energy $\Phi$ of the inherent structure $e_{IS}$ in units
of $V_0$ Vs $u/w$; $1000$ noise realizations have been
considered ($N=10$).\label{figeisN10}}
\end{figure}
\begin{figure}
\includegraphics[width=8.3cm]{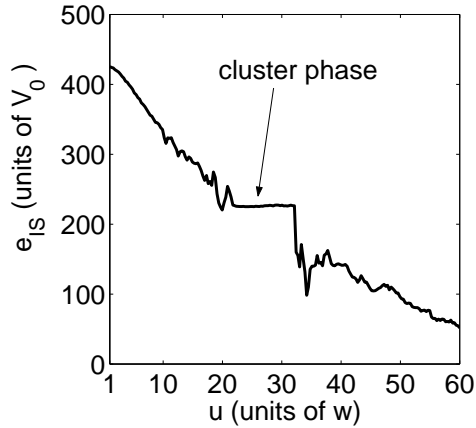}
\caption{
Average potential energy $\Phi$ of the inherent structure $e_{IS}$ in units
of $V_0$ Vs $u/w$; $100$ noise realizations have been
considered ($N=30$). \label{figeisN30}}
\end{figure}

According  to  the literature about  complex media,  \cite{Debenedetti01,Grigera03} the average  potential energy
     $e_{IS}$ of the IS is an appropriate control parameter for the dynamic phase-transition. 
In figures \ref{figeisN10} and \ref{figeisN30}, $e_{IS}$ (in units of  $V_0$) is plotted  Vs  $u/w$.
The minimum for the potential energy is obtained at large $u/w$ when all 
the solitons are in the same position and corresponds to $\Phi=0$
(condensed phase);
conversely when the solitons are uniformly distributed (small $u/w$) $\Phi$ is at maximum. 
Hence, while increasing $u/w$, $\Phi$ is reduced, due to the coagulation
mechanism.
The data in figures \ref{figeisN10} and \ref{figeisN30} show a
decrease of $e_{IS}$ Vs $u/w$ up to the first plateau, at the formation of
two clusters. 
The corresponding value of $e_{IS}/V_0$ is obtained after (\ref{PEL}),
by observing that, when two clusters of $N/2$ SSs are formed 
(for $N$ even), $\Phi/V_0\rightarrow N^2/4$ as their mutual distance
     goes to infinity. 
The plateau for $N=10$ corresponds to
$e_{IS}/V_0=25$, and to $e_{IS}/V_0=225$ for $N=30$.

The trend is conserved for various numbers of filaments. The scale on
 the $u/w$ axis changes with $N$ (see Figs. \ref{figeisN10} and
 \ref{figeisN30}) because the degree of nonlocality needed for the
phase transition obviously increases with the number of filaments.
\section{The generalized inherent structure}
For the  definitive settlement of  the phonon-saddle transition, the
so-called 
\textit{generalized inherent structure} (GIS) must  also be taken into
     account. \cite{Angelani00,Broderix00,Cavagna01,Parisi03a}
It is defined as the
     nearest stationary point of  the PEL (where all the forces
     are zero) to the final configuration. 
The latter is used as a guess in a nonlinear solver (I used the
     c05pbf NAG routine, Mark 19) for the 
$N$ equations $\partial \Phi/\partial x_p=0$, whose solution, given by
     a vector $(x_1,x_2,...,x_N)$ is just
     the GIS.
The saddle-order $K_{GIS}$ of the GIS is the number
     of  the  negative  eigenvalues  (imaginary  frequencies)  of  the
     Hessian  of $\Phi$, [see Eq. (\ref{Hessian}) below]  calculated at the GIS;  if $K_{GIS}=0$  the GIS  is a
     minimum. 

For any realization of the system there is one IS and one
     GIS. In a phonon-dominated phase the two structures are the
     same and  $K_{GIS}=0$. 
Conversely,  in a saddle-dominated phase $K_{GIS}>0$, but it
     tends  approximately   to  zero (on average over many realizations)
     in correspondence of     the     dynamic
     phase-transition. \cite{Angelani00,Broderix00,Cavagna01,Parisi03a}

To understand the physical meaning of the GIS, consider the 
principal dynamic phase transition.  While the IS corresponds to the filaments equally
distributed between the two clusters, the GIS differs
for some of the SSs positioned at intermediate places; it denotes the way the system may
escape from the energy minimum.
Clearly, if the noise-averaged $K_{GIS}$ is high the probability to find a direction in PEL to
get out from the local minimum of $\Phi$ is high: 
it somehow measures the number of escape directions from the PEL minimum. 
Actually, the average $K_{GIS}$ never reaches the zero as discussed in
\cite{Parisi03a}, because this would correspond to a complete freezing 
of the system.

\begin{figure}
\includegraphics[width=8.3cm]{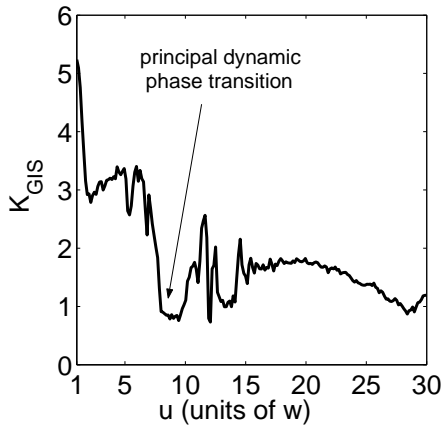}
\caption{Average saddle-order for the case $N=10$, 
other parameters as in fig. \ref{figeisN10}. \label{figKGISN10}}
\end{figure}

\begin{figure}
\includegraphics[width=8.3cm]{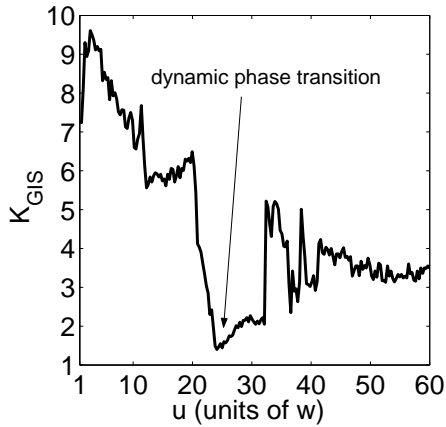}
\caption{Average saddle-order for the case $N=30$, 
other parameters as in fig. \ref{figeisN30}. \label{figKGISN30}}
\end{figure}

Figure \ref{figKGISN10} shows the noise-averaged $K_{GIS}$ Vs $u/w$ when $N=10$.
It clearly reveals a  saddle-phonon transition in proximity of
 $u/w=8$; consistently with the phase-diagram in figure
 \ref{figeisN10}. For larger values of $u/w$, the additional 
dynamic phase transitions are
 not well defined (i.e. $K_{GIS}$ stays around one or two units) due to the limited number of particles.
This also clarifies the reason for introducing a
 ``principal'' dynamic phase transition, as done above.
The latter, in the considered numerical simulations, always corresponds
to the formation of two clusters. 
This is confirmed in the case $N=30$ (shown in figure
\ref{figKGISN30}) where, due to an increased number of degrees of 
freedom,  the transition is more evident, and happens, as before, when
two clusters are generated. The transition becomes more evident as larger values of $N$ are considered, 
and it is found up to the largest considered value (i.e. $N=100$). 
Nevertheless, the case $N=10$ shows evidence of this phenomenon, which is hence observable even with a limited number
of filaments. Roughly speaking a given ensemble of filaments can self-organize in various way in order to form clusters. Two
symmetrical clusters is obviously a strong ``attractor'' for the system, because there is only one way to organize it
(conversely a larger number of clusters can be formed with different aggregations of solitons).
This mechanism strongly resembles the cluster formation in coupled chaotic maps, a well known example of complex system. 
\cite{KanekoBook}
\section{Noise quenching}
\begin{figure}
\includegraphics[width=8.3cm]{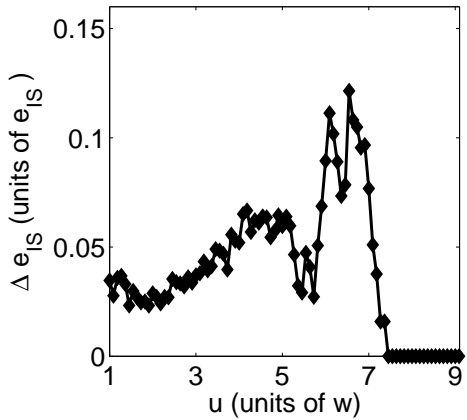}
\caption{
Standard deviation of the energy of the inherent structures for $100$ noise
realizations Vs $u/w$ ($N=10$, $t_{max}=1000$).
\label{stdisn10}}
\end{figure}

\begin{figure}
\includegraphics[width=8.3cm]{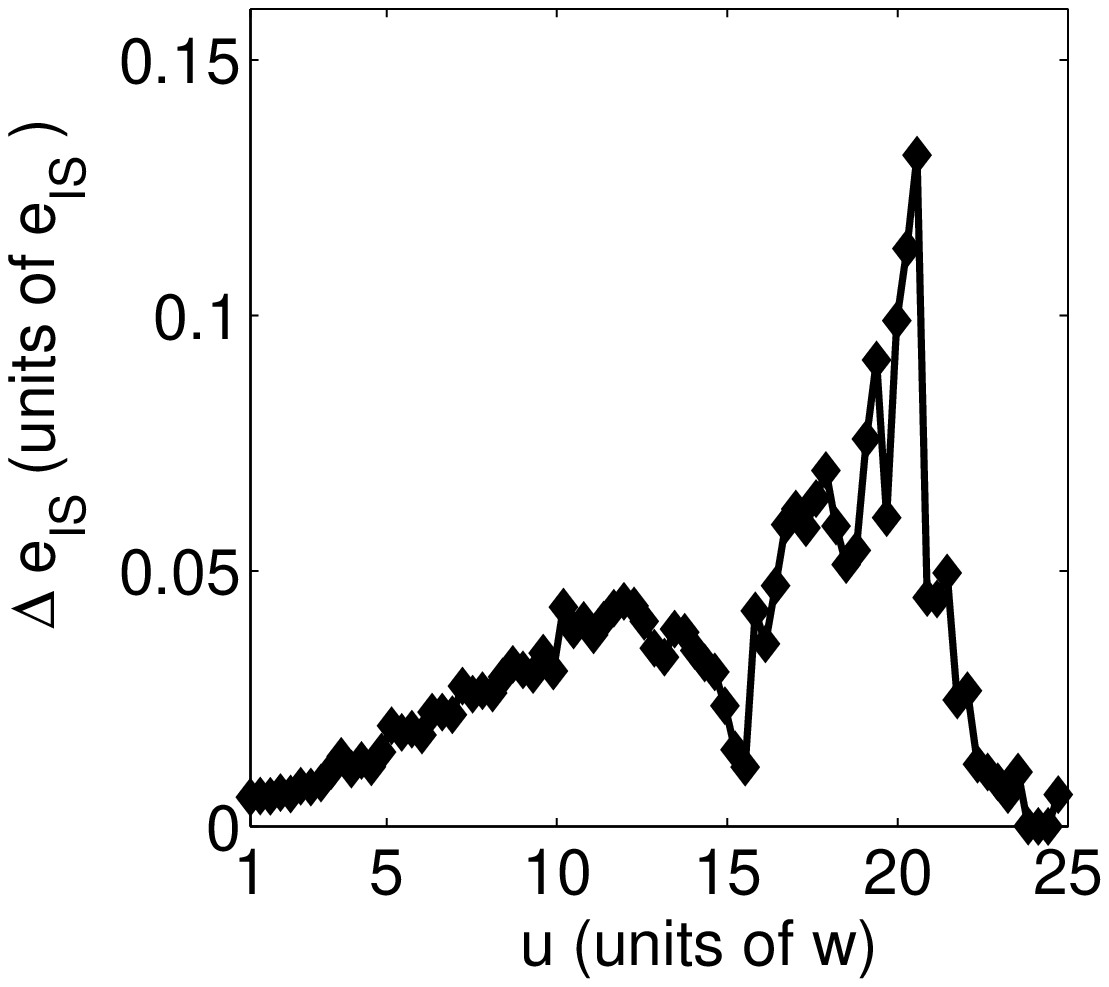}
\caption{
Standard deviation of the energy of the inherent structures for $100$ noise
realizations Vs $u/w$ ($N=30$, $t_{max}=6000$).
\label{stdisn30}}
\end{figure}

The standard deviation  $\Delta e_{IS}$ of $e_{IS}$ has a universal trend (with respect
     to number of filaments, their  initial distance and the amount of
     noise),  which is  shown in  figures \ref{stdisn10} and \ref{stdisn30}.  Moving towards a
     non-local  region  (i.e. increasing the interaction range $u/w$)
     $\Delta e_{IS}$ grows (in the landscape dominated phase \cite{Debenedetti01}) 
up  to the dynamic phase-transition, where  small values are
     again achieved (in the cluster phase).  This quantity has the same
     trend of the corresponding one investigated, for example, in \cite{Sciortino99}
     in a  Lennard-Jones material  glass.  Very  similar results  are
     obtained  when  simulating  a   larger  number  of  filaments,  and for various  values of the
     noise power.   

\begin{figure}
\includegraphics[width=8.3cm]{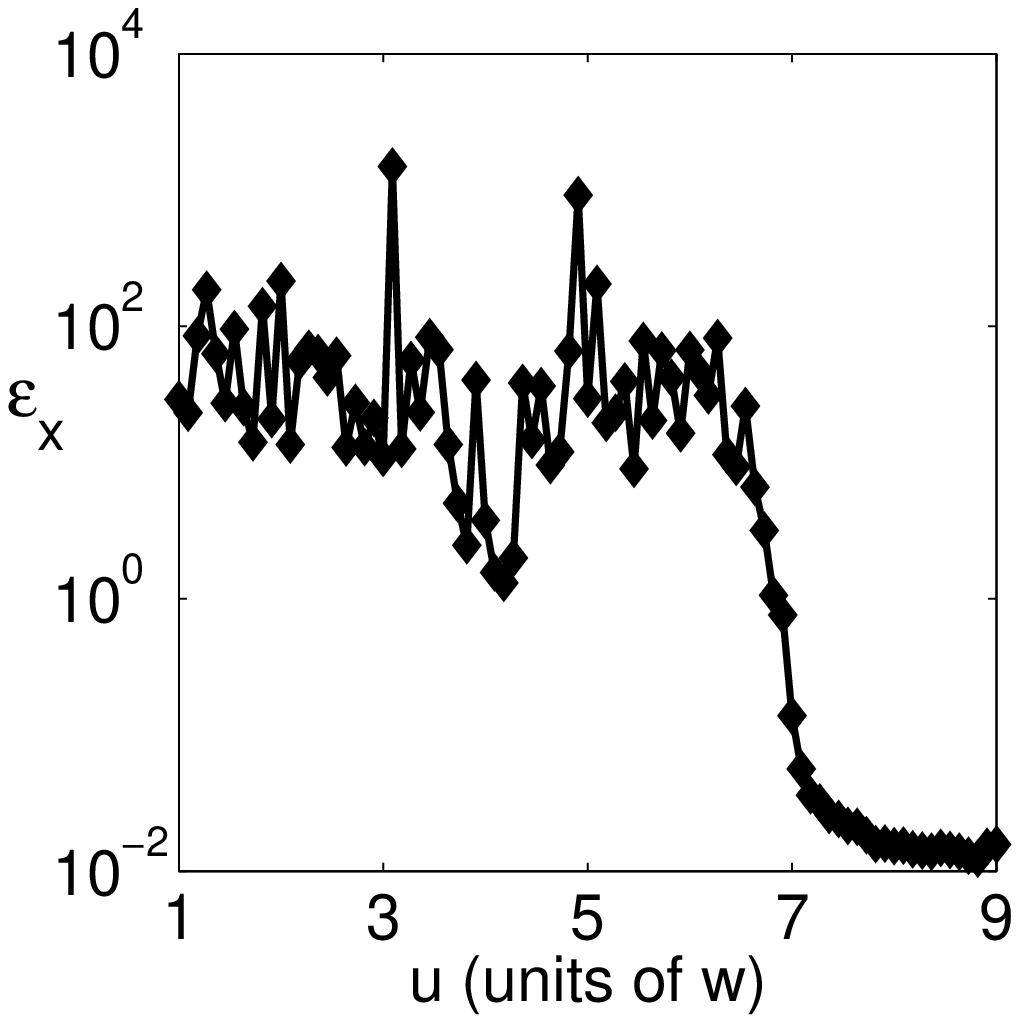}
\caption{Maximum relative deviation from the average position,
averaged over all the $N=10$ filaments Vs $u/w$.
Parameters as in fig. \ref{stdisn10}. \label{figdevposN10}}
\end{figure}

\begin{figure}
\includegraphics[width=8.3cm]{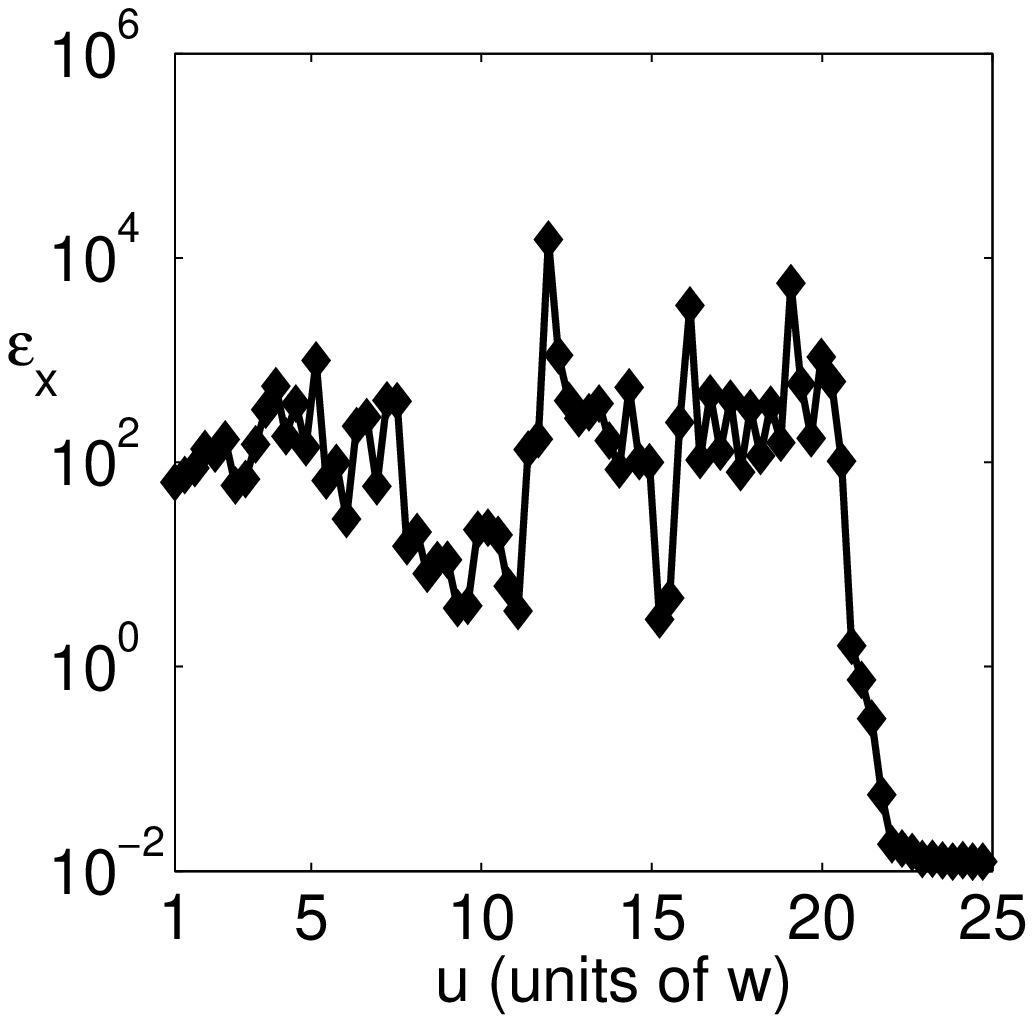}
\caption{Maximum relative deviation from the average position,
averaged over all the $N=30$ filaments Vs $u/w$.
Parameters as in fig. \ref{stdisn30}. \label{figdevposN30}}
\end{figure}

A reduction of the noise (``quenching'') in correspondence  of the  glassy-phase is hence  evident. To  confirm this
     effect, I show in figures \ref{figdevposN10} and \ref{figdevposN30}, the relative maximum deviation 
from the average position. 
This quantity, denoted $\varepsilon_x$ is calculated by
     taking the maximum deviation from the average (over the
     considered noise realizations) position $\langle x\rangle$ for each
     filament (at $t=t_{max}$), dividing by $\langle x\rangle$ and
     then averaging the resulting quantity over all the
     $N$ filaments. It measures the noise in the SS positions, 
and reproduces the same trend of $\Delta e_{IS}$.
Before the phase transition all the SSs diffuse into a wide region,
     while after the phase transition they are locked inside each
     cluster.

In the experiments this phenomenon is resolved in time, while changing the control parameter for 
the nonlocality (e.g. the voltage in NLC experiments \cite{Peccianti05}). 
This means that,
     since the medium is fluctuating, before the transition the number
     and the positions of the clusters are rapidly varying. Conversely,
     when the two clusters are formed, the intensity profile ``slows
     down'', and the noise is quenched; this resembles the ``critical slowing
     down''\cite{PhysicaD} in glassy material system 
(see also the section VIII and the movie cited below).
\section{Vibrational spectra and the random matrices}
A variety of issues spontaneously rises,
once some analogy with a disordered medium has been ascertained.
In particular, those concerning the spectrum of the fluctuations
and ultimately the propagation of ``sound-waves'' (or better 
``displacement-waves''). These are associated to the vibrational 
spectra of the ensemble of solitons.

The analysis of sound-waves is
one of the most important issues in the physics of glassy
systems (see for example \cite{Scopigno04} and references therein).
According to some authors, the appearance of ultra-high frequency
sound can be related to the vibrational spectrum of the material and,
in particular,
and to an excess of states denoted ``boson peak'' (see \cite{Parisi03}
and references therein).
The transposition of these ideas to nonlinear optical propagation is
beyond the scope of this article. However, it interesting to observe
that one of the most successful theories of structural glasses,
the random-matrices approach, \cite{Grigera01,Grigera03} also seems well suited to describe
the vibrational spectra of the positions of a number of optical
spatial solitons in a disordered medium.

Given some configuration of the filaments $(x_1,x_2,...,x_N)$, which can be
either the instantaneous distribution at $t_{max}$, or the IS, \cite{Parisi03}
the vibrational spectra can be found as the eigenvalues of the
Hessian matrix $H_{pq}$ ($p,q=1...N$):
\begin{equation}
\label{Hessian}
H_{pq}=\frac{\partial^2 \Phi}{\partial x_p \partial
x_q}=\delta_{pq}\displaystyle
\sum_{k=1}^{N} V''(x_p-x_k)-V''(x_p-x_q)\text{,}
\end{equation}
with $V''\equiv d^2 V/dx^2$.
Due to noise (and eventually to chaos), the considered configuration
has a certain statistical distribution; the problem is hence
reduced to find the corresponding statistical distribution of the eigenvalues of $H_{pq}$.
Various approaches have been developed and successfully applied to
explain some material glass features. \cite{Grigera01,Grigera03} What follows suggests that the 
random matrices approach could be very fruitful even in this field of research.

Consider an experiment in which the position of each SS can be
retrieved by the scattered light from the top of the sample (as those in the
mentioned literature on \textit{nematicons} \cite{Assanto03,Peccianti04nature}); 
the resulting images appear as a number of superimposed 
light filaments, each with its own few-microns waist, with 
overall intensity distribution $I(x,z)$.
If a soliton profile $I_S$ is associated to each SS, as
discussed above, $I(x,z)$ can be interpreted as a coarse-grained 
density of particles evolving along $z$.
The squared modulus of the double-Fourier transform of the
image $I(x,z)$, denoted
$S(k_x,k_z)$ (averaged over a given number of noise realizations)
 can be interpreted as the so-called ``dynamic structure
factor'' \cite{SimpleLiquids} of the soft-medium realized by the SSs, which play the role of interacting Brownian molecules. 
$S(k_x,k_z)$ gives the frequency content in $k_z$
of the $z$-evolution of the spatial ``mode'' at
$k_x$.
In other words, once fixed $k_x$, $S(k_x,k_z)$ provides information on
the dynamics of each intensity perturbation with period $2\pi/k_x$.
\footnote{
This is rigorously true only after the transient (along
$z$) during which the clusters are formed. However in the following 
the overall dynamics from $t=0$ is considered, for simplicity sake. I
have checked that no substantial differences arise in the spectrum
if the transient is removed before making the Fourier transform of $I(x,z)$. 
}

\begin{figure}
\includegraphics[width=8.3cm]{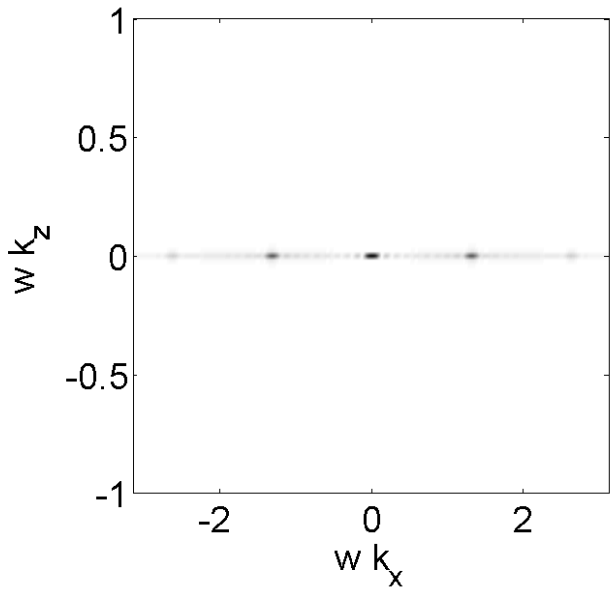}
\caption{
Pseudo-color plot of the squared modulus of the Fourier transform of the course grained density 
(overall intensity profile)when $N=10$, $u/w=1$ and $t_{max}=200$,
averaged over $100$ noise realizations. \label{RMu1}}
\end{figure}

Consider, for example, the case in which all the solitons travel
approximately parallel, with a reduced interaction.
In this case, if $\Delta x$ is the average mutual distance,
$S(k_x,k_z)$ is expected to be approximately given by a series
of  peaks around $k_x=2\pi m/\Delta x$ and $k_z=0$, with $m=0,1,2,...$.
That is the $k_z$-bandwidth of each ``mode'' at $k_x$ is very small.
This happens, for example in the case $N=10$, 
when $u/w=1$, as shown in the inset in figure \ref{figN10}.
Taking the
corresponding numerical solution of eqs. (\ref{particle3}), associating to each
trajectory a soliton profile (a Gaussian profile in units such that
$w=1$), evaluating the squared modulus of the Fourier transform of the
resulting $I(x,z)$, and finally averaging over a given number of
noise realizations, Fig. \ref{RMu1} is obtained, which appears as anticipated.

\begin{figure}
\includegraphics[width=8.3cm]{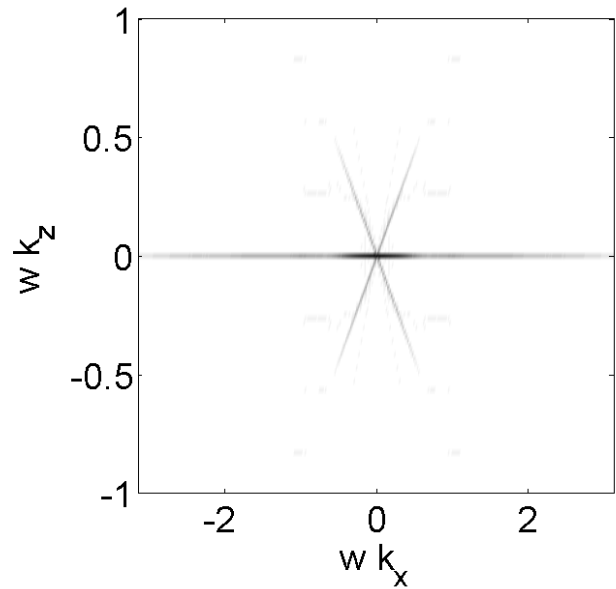}
\caption{
Pseudo-color plot of the squared modulus of the Fourier transform of the course grained density 
(overall intensity profile)when $N=10$, $u/w=8$ and $t_{max}=200$,
averaged over $100$ noise realizations. \label{RMu8}}
\end{figure}

The aim of the random matrices approach is to predict the 
shape of $S(k_x,k_z)$ with respect to some control parameter. 
In particular, under very general hypotheses, 
it has been shown that, when the interaction range grows (or
equivalently the density of the particles is increased), $S(k_x,k_z)$
develops a Brillouin peak with position that is linearly dependent on $k_x$.\cite{Grigera01}
This corresponds to an X-shape in the two dimensional level-plot of
$S(k_x,k_z)$.  Repeating the analysis of
figure \ref{RMu1}, for $u/w$ at the dynamic phase transition,
should provide some evidence of the predicted X-shape.
This is exactly what happens, as shown in figure \ref{RMu8}. 

The formation of a Brillouin peak is obviously related 
to the geometry of the filaments distribution. 
However, it can be also associated to a movement of energy (or
equivalently of ``mass'', which corresponds to the beam power into each
filament, as discussed above).
Indeed, when two clusters are formed and move
far apart each other, there is 
an evident energy transfer along $x$. In some 
sense the cluster movement is related to some wave-packet motion, in 
perfect analogy with sound waves in disordered media.
\section{Numerical simulations of a stochastic PDE model}
In this section I consider a specific example in order to validate the previous arguments.
Some numerical simulations of a stochastic PDE system (which is well known in the deterministic limit and has 
been successfully compared with experiments)
 are reported and reproduce the features previously described by the Brownian dynamics approach.
For the sake of clearness and compactness, I will show sample results for nonlocalities up to the first dynamic transition (i.e. the formation
of two dominant clusters). Additional data will be reported elsewhere.

The model equations are those of the so-called 1+1D exponential non-locality, which 
well describes solitons in liquid crystals \cite{Conti03, Peccianti03}, as well as thermal nonlinearities  
and plasmas \cite{Litvak75,Yakimenko04}, and has been thoroughly studied (in absence of noise) 
in the literature (see e.g. \cite{Bang02, Krolikowski04} and references therein). Using adimensional variables the PDE system 
(corresponding to Eq. (\ref{Foch}))
reads as
\begin{equation}
\label{PDE1}
\begin{array}{l}
i \partial_\zeta \psi+\partial_{\xi \xi} \psi +\rho \psi=0\\
-\sigma^2 \partial_{\xi \xi} \rho +\rho=|\psi|^2 +A \eta(\xi,\zeta)\text{.}
\end{array}
\end{equation}
In (\ref{PDE1}) $\xi$ is the normalized transverse coordinate, $\zeta$ is the normalized propagation distance, $\psi$ is a complex field, corresponding to the electromagnetic field; $\rho$ is 
the medium disturbance, which can be the director angle of NLC, or the temperature, or the density of the medium, depending
on the specific physical system;  its fluctuations are taken into account by a Langevin term, which is a white Gaussian
stochastic process, such that $\langle \eta(\xi,\zeta) \eta(\xi',\zeta')\rangle=\delta(\xi-\xi')\delta(\zeta-\zeta')$, and $A$ measures the amount of noise.
$\sigma^2$ is takes into account the nonlocality; as $\sigma^2=0$, the model reduces to the local integrable nonlinear
Schr\"odinger equation; the degree of nonlocality increases with as $\sigma^2$. 
Equations (\ref{PDE1}) 
are solved by a pseudospectral approach (see e.g. \cite{BoydSpectralBook}), which maps (\ref{PDE1}), via discrete Fourier transform, into 
coupled stochastic ordinary equations, which are then solved by the Heun algorithm. \cite{Greiner88}
\begin{figure}
\includegraphics[width=8.3cm]{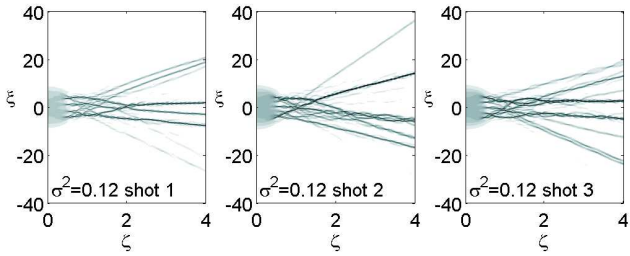}
\caption{
Three realizations of the numerical solution of the 
stochastic PDE (\ref{PDE1}) for $\sigma^2=0.12$.
\label{figurePDE1}}
\end{figure}
\begin{figure}
\includegraphics[width=8.3cm]{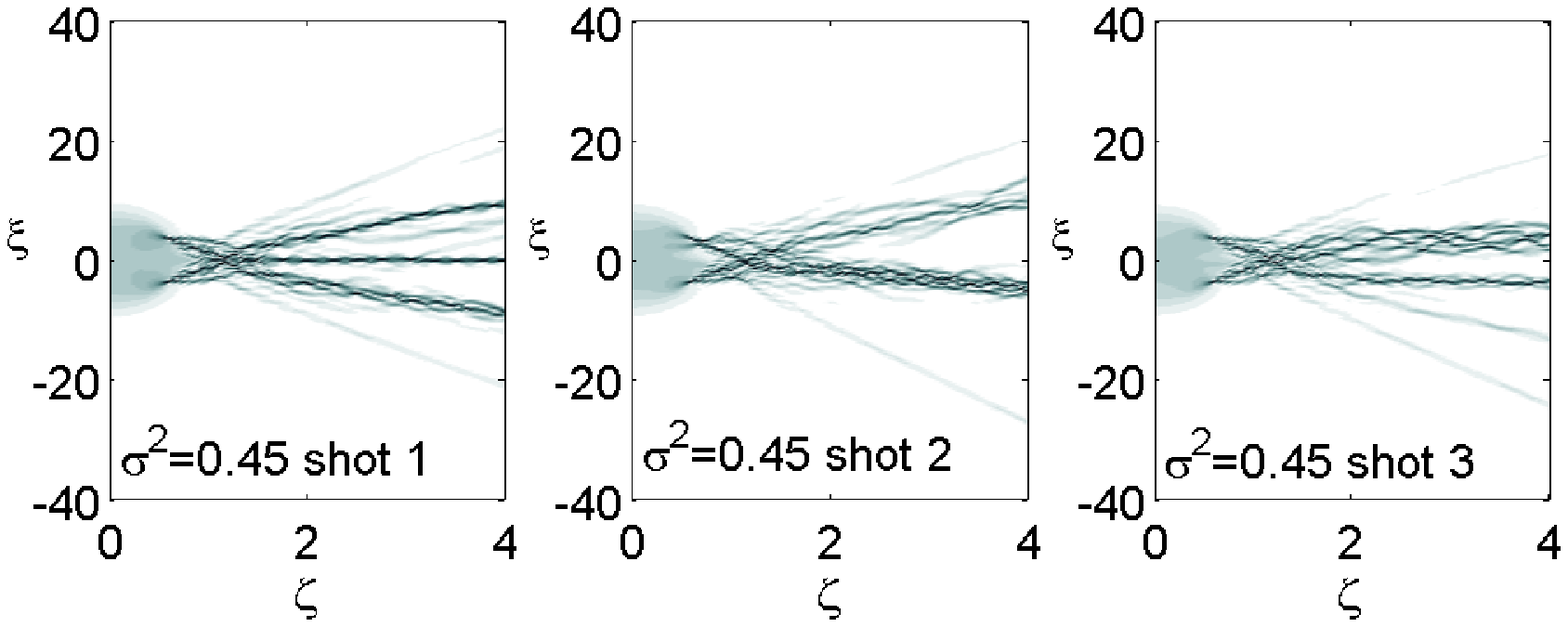}
\caption{
Three realizations of the numerical solution of the 
stochastic PDE (\ref{PDE1}) for $\sigma^2=0.45$.
\label{figurePDE2}}
\end{figure}
\begin{figure}
\includegraphics[width=8.3cm]{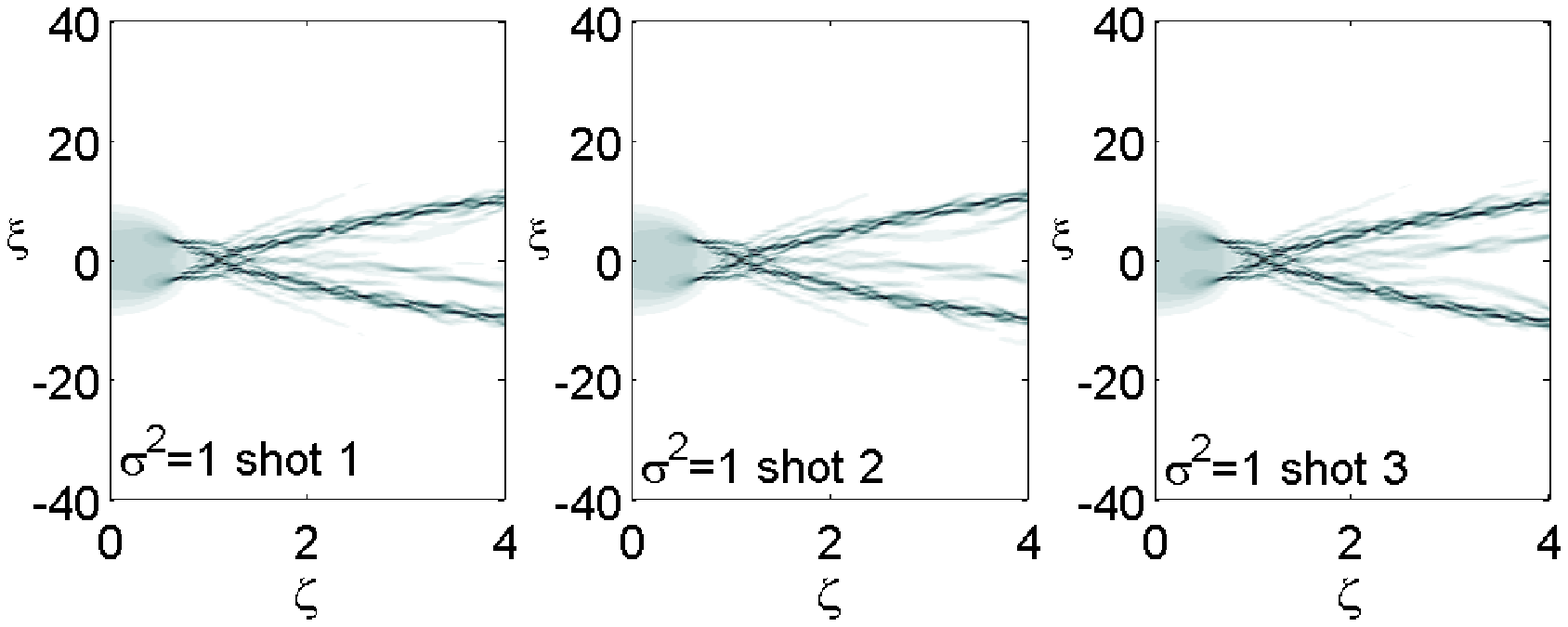}
\caption{
Three realizations of the numerical solution of the 
stochastic PDE (\ref{PDE1}) for $\sigma^2=1$.
\label{figurePDE3}}
\end{figure}
\begin{figure}
\includegraphics[width=6cm]{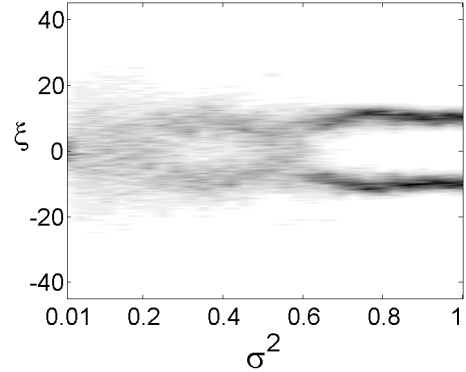}
\caption{
Average intensity distribution at $\zeta=4$ over $100$ noise realizations as a function
of $\xi$ and $\sigma^2$.
\label{figurePDEmedie}}
\end{figure}
\begin{figure}
\includegraphics[width=6cm]{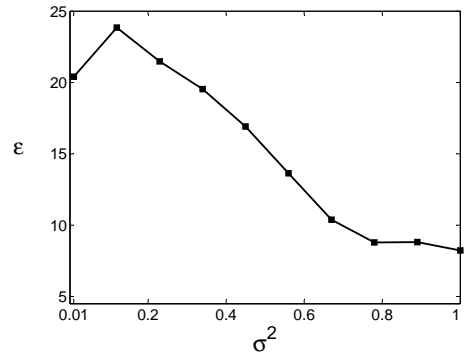}
\caption{
Maximum relative deviation from the average profile versus the nonlocality parameter calculated at $\zeta=4$.
\label{figurePDEerr}}
\end{figure}
\begin{figure}
\includegraphics[width=6cm]{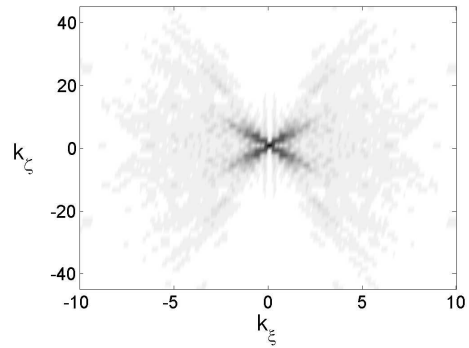}
\caption{
Two dimensional Fourier transform of the intensity profile, average over $100$ realization, when $\sigma^2=1$.
\label{figurePDEspectra1}}
\end{figure}

In figures \ref{figurePDE1},\ref{figurePDE2} and \ref{figurePDE3} I show some realizations obtained with different values of the nonlocality parameters $\sigma^2$.
The input profile is taken to be a super-Gaussian:
\begin{equation}
\psi(\xi,0)=\exp(-(\xi/10)^4)\text{,}
\end{equation}
such that it corresponds to a flat intensity profile, on which various solitons are generated via modulational instability and
interact while propagating along the $\zeta$ direction.
(The reported results correspond to $A=10^{-3}$).

Consistently with the analysis reported above, for small nonlocality various solitons are generated and travel almost independently, 
while at higher nonlocality clusters are formed,
and two dominant aggregates of solitons are clearly evident when $\sigma^2=1$.

In figure \ref{figurePDEmedie} 
I show the intensity profile at $\zeta=4$ averaged over $100$ realizations, it clearly reproduces the features 
in figures \ref{figmeanposn10} and  \ref{figmeanposn30},
where two dominant clusters appear at a threshold value of nonlocality.

Note from figs. \ref{figurePDE1}-\ref{figurePDE3}
 that, for small $\sigma^2$, the various shots are very different from each other,
while at high nonlocality the two clusters distribution is ``freezed.''
\footnote{A movie reporting various noise realizations for different $\sigma^2$ can be downloaded at \url{http://nlo.phys.uniroma1.it/complexity.htm}}

The noise quenching mechanism is quantified in figure \ref{figurePDEerr}, where I 
show the noise figure $\varepsilon=\varepsilon(\zeta=4)$,  determined as
\begin{equation}
\varepsilon(\zeta)=\langle max_{\xi}\left(\frac{|\psi(\xi,\zeta)|^2}{\langle |\psi(\xi,\zeta)|^2 \rangle}-1\right)\rangle\text{,}
\end{equation}
and corresponding to  the maximum (along the $\xi$ direction) relative intensity deviation with respect to the average profile at $\zeta=4$. 
Figure \ref{figurePDEerr} shows a drastic reduction of noise at high nonlocality, in correspondence
of the two clusters. 

Finally, the vibrational spectrum (as described in the previous section) is calculated:
the intensity distribution $|\psi(\xi,\zeta)|^2$, for a fixed $\sigma^2$, is 
Fourier transformed and averaged over the considered $100$ realizations. 
The result for $\sigma^2=1$ is shown in figure \ref{figurePDEspectra1} and clearly displays the X-shape,
addressed above. A similar picture is obtained for other values of nonlocality in correspondence
of the two clusters formation. 
\section{Conclusion}
The basic aim of this manuscript is to point out one of the possible connections between
the physics of complex media and that of intense laser light
interacting with matter. 
During nonlinear optical phenomena in disordered media (or even
in media where the disorder is induced by the input laser beam), 
the behavior of light can be interpreted using the same paradigms of
modern statistical physics. 

If a light filament, or spatial optical soliton, can be treated as
a classical particle, many interacting spatial solitons correspond to
a liquid or high density gas. If some noise is present in the system, a
Brownian dynamics model is readily introduced, and the filaments
behave like particles dispersed into a solvent, which is one of the
simplest definitions of soft-matter.
The dynamic phase transition is the natural way to describe 
phenomenological transformations, like the formation of clusters.
This has been shown by numerical experiments in this article and
experimentally in future publications.

Nonlinear optics can be hence used to test, theoretically and
experimentally, some of the ideas of the physics of complexity, 
as well as the latter can be used to explain many high-field 
phenomena, like laser-filaments generation and interaction.
In this manuscript these ideas have been applied to explain a possible 
manifestation of complex light.
\acknowledgements{
It is my pleasure to thank L. Angelani, B. Crosignani, E. Del Re, G. Ruocco, F. Sciortino and 
S. Trillo for their interest and for many stimulating discussions.
A particular acknowledgement goes to M. Peccianti and G. Assanto, 
who made possible the
experimental investigation of dynamic phase transitions of light,
as well as a very prolific research period for the author.}


\end{document}